\documentclass[12pt,a4paper]{article}
\pdfoutput=1
\usepackage[utf8]{inputenc}
\usepackage{epsf,amsmath,amssymb,graphicx,dcolumn}
\usepackage{caption}
\usepackage{subcaption}
\usepackage{scalefnt,ulem,pstricks}
\usepackage{booktabs,multirow,tabularx}
\usepackage{cite,hyperref}
\usepackage{cleveref}
\usepackage{color}
\usepackage{rotating}
\usepackage{microtype}
\usepackage[titletoc,title]{appendix}
\usepackage[numbers,sort&compress]{natbib}
 
\newcommand{\sphid}[1]{}

\providecommand{\href}[2]{#2}

\newcommand\as{\alpha_{\mathrm{S}}}

\def\to{\rightarrow}

\def\mz{m_Z}

\newcommand\powheg{{\sc Powheg}}
\newcommand\Matrix{{\sc Matrix}}
\newcommand\Munich{{\sc Munich}}
\newcommand\OpenLoops{{\sc OpenLoops}}
\newcommand\Collier{{\sc Collier}}

\newcommand{\CutTools}{{\sc CutTools}}
\newcommand{\OneLOop}{{\sc OneLOop}}
\newcommand{\qt}{\ensuremath{q_T}}
\newcommand{\pt}{\ensuremath{p_T}}

\newcommand{\fig}[1]{Figure~\ref{#1}}

\newcommand{\tab}[1]{Table~\ref{#1}}

\def\citere#1{\mbox{Ref.~\cite{#1}}}
\def\citeres#1{\mbox{Refs.~\cite{#1}}}

\newcommand{\zz}{\ensuremath{ZZ}}
\newcommand{\ww}{\ensuremath{W^+W^-}}

\newcommand{\z}{\ensuremath{Z}}

\newcommand{\abbrev}{}

\newcommand{\nlo}{\text{\abbrev NLO}}
\newcommand{\nnlo}{\text{\abbrev NNLO}}

\newcommand{\qcd}{{\abbrev QCD}}

\newcommand\Bstrut{\rule[-1.5ex]{0pt}{0pt}}   % = `bottom' strut

\newcommand{\elle}{\ensuremath{\ell}}

\newcommand{\ptwm}{\ensuremath{p_{T,\ell^-\nu_\ell}}}

\newcommand{\mll}{\ensuremath{m_{\ell^+\ell^-}}}

\newcommand{\ptl}{\ensuremath{p_{T,\ell}}}
\newcommand{\ptll}{\ensuremath{p_{T,\ell\ell}}}

\newcommand{\ptlone}{\ensuremath{p_{T,\ell_1}}}

\newcommand{\ptmiss}{\ensuremath{p_{T}^{\text{miss}}}}
\newcommand{\dphill}{\ensuremath{\Delta\phi_{\ell\ell}}}

\newcommand{\etal}{\ensuremath{\eta_{\ell}}}

\newcommand{\dyzz}{\ensuremath{\Delta y_{Z_1,Z_2}}}

\newcommand{\dphilzlz}{\ensuremath{\Delta\phi_{\ell^+_{Z_1},\ell^-_{Z_1}}}}
\newcommand{\ptzone}{\ensuremath{p_{T,Z_1}}}
\newcommand{\njets}{\ensuremath{N_{\rm jets}}}
\newcommand{\mtzz}{\ensuremath{m_{T,ZZ}}}

\begin{document} 
\begin{flushright}
\vspace*{-1.5cm}
CERN-TH-2018-127\\
\end{flushright}
\vspace{0.cm}

\begin{center}
{\Large \bf $\zz$ production at the LHC:\\[0.2cm] NNLO predictions for $2\ell2\nu$ and $4\ell$ signatures}
\end{center}

\begin{center}
{\bf Stefan Kallweit} and {\bf Marius Wiesemann}

TH Division, Physics Department, CERN, CH-1211 Geneva 23, Switzerland

\href{mailto:stefan.kallweit@cern.ch}{\tt stefan.kallweit@cern.ch}\\
\href{mailto:marius.wiesemann@cern.ch}{\tt marius.wiesemann@cern.ch}

\end{center}

\begin{center} {\bf Abstract} \end{center}\vspace{-1cm}
\begin{quote}
\pretolerance 10000

We consider QCD radiative corrections to \zz{} production for all experimentally 
relevant leptonic processes.
We report on a novel computation of next-to-next-to-leading-order (NNLO) corrections to
the diboson signature with two charged leptons and missing transverse energy 
($\ell\ell$+$E_T^{\rm miss}$). 
All relevant final states are considered: $\ell\ell\nu_\ell\nu_\ell$,  $\ell\ell\nu_{\ell'}\nu_{\ell'}$ and $\ell\nu_{\ell}\ell'\nu_{\ell'}$.
We also study processes with four charged leptons: $\ell\ell\ell\ell$ and $\ell\ell\ell'\ell'$.
For the first time NNLO accuracy is achieved for a  
process mixing two double-resonant diboson topologies
(\zz{}/\ww{}$\to\ell\ell\nu_\ell\nu_\ell$).
We find good agreement 
with ATLAS data at $8$\,TeV.
NNLO corrections are large ($5$--$20$\% and more), and interference effects between \zz{} 
and \ww{} resonances turn out to be negligible in most cases.
\end{quote}

\parskip = 1.2ex 

Diboson processes play a major role in the rich physics programme of the LHC. 
The intriguing nature of these processes combined with their rather clean experimental 
signatures and relatively large cross sections render them ideal for Standard Model (SM) 
precision measurements. The precise knowledge of diboson rates and distributions provides
a strong test of the gauge-symmetry structure of electroweak (EW) interactions and 
the mechanism of EW symmetry breaking. They also serve as important probes of new 
physics phenomena in direct and indirect searches. Diboson final states, in 
particular \zz{} and \ww{}, are also extensively used in Higgs-boson measurements.

The production of \zz{} pairs yields the smallest cross section among the diboson processes. 
Nevertheless, its pure experimental signature with four charged leptons in the final state 
facilitates a clean measurement so that it has already been used in 
a combination of ATLAS and CMS data to constrain 
anomalous trilinear gauge couplings~\cite{ATLAS:2016hao}. 
\zz{} production at the LHC has been measured 
at 7\,TeV~\cite{Aad:2011xj,Chatrchyan:2012sga,Aad:2012awa},
8\,TeV~\cite{Chatrchyan:2013oev,CMS:2014xja,Khachatryan:2015pba,Aad:2015rka,Aaboud:2016urj}, 
and 13\,TeV~\cite{Aad:2015zqe,Khachatryan:2016txa,Aaboud:2017rwm,Sirunyan:2017zjc}. Also searches for new heavy \zz{} resonances involving both 
charged leptons and neutrinos have been performed, see \citere{Aaboud:2017rel} for example.

Theoretical predictions for \zz{} production at next-to-leading order (\nlo{}) \qcd{} 
were obtained a long time ago for both on-shell \z{} bosons~\cite{Ohnemus:1990za,Mele:1990bq} and 
their fully leptonic final states \cite{Ohnemus:1994ff,Campbell:1999ah,Dixon:1999di,Dixon:1998py}. 
Perturbative corrections beyond NLO QCD are indispensable to reach the precision demanded by present \zz{} measurements. 
NLO EW corrections are known for stable \z{} bosons \cite{Accomando:2004de,Bierweiler:2013dja,Baglio:2013toa} 
and including their full off-shell treatment for leptonic final states~\cite{Biedermann:2016yvs,Biedermann:2016lvg,Kallweit:2017khh}.
$\zz$+${\rm jet}$ production was computed at NLO QCD \cite{Binoth:2009wk}.
The loop-induced $gg\to\zz+X$ subprocess, which provides a separately finite ${\cal O}(\as^2)$ contribution, is known at leading order (LO) \cite{Glover:1988rg,Dicus:1987dj,Matsuura:1991pj,Zecher:1994kb,Binoth:2008pr,Campbell:2011bn,Kauer:2013qba,Cascioli:2013gfa,Campbell:2013una,Kauer:2015dma} and was recently computed at NLO considering only 
$gg$-initiated partonic channels \cite{Caola:2015psa,Caola:2016trd,Alioli:2016xab}, using the two-loop helicity amplitudes for $gg\to VV'$ of \citeres{Caola:2015ila,vonManteuffel:2015msa}. 
\nnlo{} \qcd{} corrections to on-shell \zz{} production were first evaluated 
in \citere{Cascioli:2014yka}, and later in \citere{Heinrich:2017bvg}. Using the
two-loop helicity amplitudes for $q\bar{q}\to VV'$~\cite{Gehrmann:2014bfa,Caola:2014iua,Gehrmann:2015ora}, differential predictions in the
four-lepton channels ($\ell\ell\ell\ell$ and $\ell\ell\ell'\ell'$) were presented 
in \citere{Grazzini:2015hta}.

In this paper we complete \nnlo{} QCD corrections to \zz{} production by considering all 
experimentally relevant leptonic final states. 
Our computations are fully differential in the momenta of the final-state leptons, and we
account for off-shell effects and spin correlations 
by consistently including all resonant and non-resonant topologies.
For the first time, we obtain \nnlo{}-accurate predictions for the (same-flavour) dilepton plus missing transverse energy signature ($\ell\ell$+$E_T^{\rm miss}$),
which involves all processes with two opposite-charge leptons and two neutrinos in the final state ($\ell\ell\nu_\ell\nu_\ell$,  $\ell\ell\nu_{\ell'}\nu_{\ell'}$ and 
$\ell\nu_{\ell}\ell'\nu_{\ell'}$). The process $\ell\ell\nu_\ell\nu_\ell$ is particularly interesting as it mixes \zz{} and \ww{} topologies, which will be 
studied in detail. For completeness we also compute \nnlo{} corrections to the four-lepton channels ($\ell\ell\ell\ell$ and $\ell\ell\ell'\ell'$).
Phenomenological predictions at \nnlo{} for all of the aforementioned leptonic processes are compared to LHC data at 8\,TeV.

We employ the computational framework \Matrix{} \cite{Grazzini:2017mhc}.
All tree-level and one-loop amplitudes are evaluated with 
\OpenLoops{}\footnote{\OpenLoops{} relies on the fast and stable
tensor reduction of \Collier{}~\cite{Denner:2014gla,Denner:2016kdg},
supported by a rescue system based on quad-precision
\CutTools\cite{Ossola:2007ax} with \OneLOop\cite{vanHameren:2010cp}
to deal with exceptional phase-space
points.}~\cite{Cascioli:2011va,Buccioni:2017yxi}.
At two-loop level we use the $q\bar{q}\to VV'$ amplitudes of \citere{Gehrmann:2015ora},
and implement the leptonic final states with two charged leptons and two neutrinos 
as well as with four charged leptons. NNLO accuracy is achieved  
by a fully general implementation of the \qt{}-subtraction formalism \cite{Catani:2007vq}
within \Matrix{}. The NLO parts therein (for \zz{} and \zz{}+$1$-jet)
are performed by \Munich{}\footnote{The Monte
Carlo program \Munich{} features a general implementation of an
efficient, multi-channel based phase-space integration and computes
both NLO QCD and NLO EW~\cite{Kallweit:2014xda,Kallweit:2015dum} corrections
to arbitrary SM processes.}~\cite{munich}, which employs the 
Catani--Seymour dipole subtraction method \cite{Catani:1996jh,Catani:1996vz}.
The \Matrix{} framework features NNLO QCD corrections to a large number of colour-singlet processes at hadron colliders, and has already been used to obtain several 
state-of-the-art NNLO predictions \cite{Grazzini:2013bna,Grazzini:2015nwa,Cascioli:2014yka,Grazzini:2015hta,Gehrmann:2014fva,Grazzini:2016ctr,Grazzini:2016swo,Grazzini:2017ckn,deFlorian:2016uhr,Grazzini:2018bsd}.\footnote{It was
  also used in the NNLL+NNLO computation of \citere{Grazzini:2015wpa}, and in the NNLOPS computation of \citere{Re:2018vac}.}
  
\begin{figure}
\begin{center}
\begin{tabular}{ccc}
\includegraphics[width=.25\textwidth]{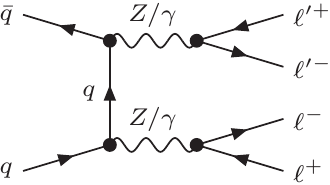} & \quad \quad \quad\quad &
\includegraphics[width=.25\textwidth]{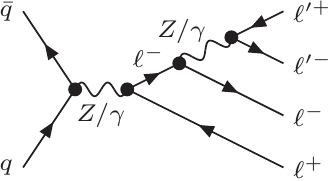} \\[0ex]
(a) & & (b)\\[-0.5ex]
\end{tabular}
\end{center}
\caption[]{\label{fig:diag}{Born-level Feynman diagrams for \zz{}
    production with four charged final-state leptons.}}
\end{figure}
  
\begin{figure}
\begin{center}
\begin{tabular}{ccc}
\includegraphics[width=.25\textwidth]{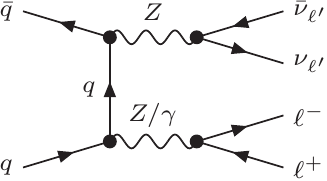} &  \quad \quad\quad&
\includegraphics[width=.25\textwidth]{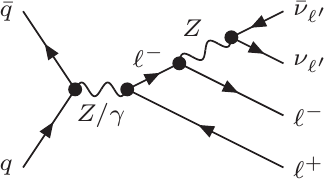} \\[0ex]
(a) & & (b)
\end{tabular}
\vspace*{.5ex}
\begin{center}
\begin{tabular}{ccccc}
\includegraphics[width=.25\textwidth]{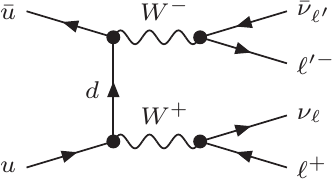} & \quad\quad&
\includegraphics[width=.25\textwidth]{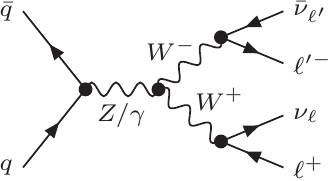} &  \quad\quad&
\includegraphics[width=.25\textwidth]{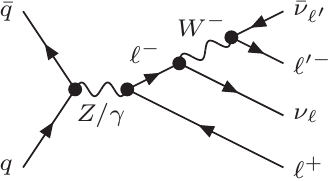} \\[0ex]
(c) & & (d) & & (e)
\end{tabular}
\end{center}
\caption[]{\label{fig:diag2}{Born-level Feynman diagrams for the production of 
two charged leptons and two neutrinos: (a-b) topologies of 
\zz{} production contributing to the process $pp\to\ell^+\ell^-\,\nu_{\ell'}\bar\nu_{\ell'}$ ($\elle\neq\elle^\prime$);
(c-e) topologies of \ww{} production contributing to the process $pp\to\ell^+\nu_{\ell}\,\ell'^-\bar\nu_{\ell'}$ ($\elle\neq\elle^\prime$); for $\elle = \elle^\prime$ all diagrams contribute to 
the process $pp\to\ell^+\ell^-\,\nu_{\ell}\bar\nu_{\ell}$, thereby mixing \zz{} and \ww{} topologies.}}
\end{center}
\end{figure}

We consider all leptonic signatures relevant for \zz{} measurements at the LHC. 
On the one hand, we compute the four-lepton ($4\ell$) processes
\begin{align}
pp \rightarrow \ell^+\ell^-\,\ell'^+\ell'^-+X,\nonumber
\end{align}
with different-flavour (DF) leptons ($\ell \neq \ell'$), denoted as $\ell\ell\ell'\ell'$,
and same-flavour (SF) leptons ($\ell = \ell'$), denoted as $\ell\ell\ell\ell$.  
Representative
LO diagrams are shown in \fig{fig:diag}. They involve both double-resonant $t$-channel 
\zz{} production~(panel a) and single-resonant $s$-channel Drell--Yan (DY) topologies~(panel b).
On the other hand, we compute processes with two charged leptons and two neutrinos ($2\ell2\nu$) in the final state,
\begin{align}
pp \rightarrow \ell^+\ell^-\,\nu_{\ell'}\bar\nu_{\ell'}+X,\quad pp \rightarrow \ell^+\nu_{\ell}\,\ell'^-\bar\nu_{\ell'}+X\text{,\quad and }\; pp \rightarrow \ell^+\ell^-\,\nu_{\ell}\bar\nu_{\ell}+X,\quad\text{ with }\ell \neq \ell'. \nonumber
\end{align}
Representative LO diagrams are shown in \fig{fig:diag2}.
In the first process the flavour of the neutrinos does not match the 
flavour of the 
charged leptons, and it features double-resonant \zz{} contributions~(panel a) as well as 
DY-type topologies~(panel b). 
In the second process the two charged leptons are of different flavours, 
and it features 
double-resonant \ww{} contributions~(panels c and d) as well as DY-type topologies~(panel e). 
In the third process all leptons and neutrinos are of the same flavour, 
and the topologies of the first two processes mix in the matrix elements.

All of the aforementioned processes with charged leptons $\ell,\ell'\in\{e,\mu\}$ and neutrinos $\nu_\ell,\nu_{\ell'}\in\{\nu_e,\nu_\mu,\nu_\tau\}$ are studied.
The loop-induced $gg$ component is part of the NNLO corrections to these processes 
and therefore included. The same is true for resonant Higgs-boson topologies, which 
also start contributing at $\mathcal{O}(\alpha_s^2)$.

A significant complication of the processes
$pp \rightarrow \ell^+\nu_{\ell}\,\ell'^-\bar\nu_{\ell'}$ and $pp \rightarrow \ell^+\ell^-\,\nu_{\ell}\bar\nu_{\ell}$
is posed by the contamination from resonant top-quark contributions
with $t\rightarrow Wb$ decays, which enters radiative corrections through diagrams featuring external bottom quarks.
In the context of \ww{} production \cite{Gehrmann:2014fva,Grazzini:2016ctr} 
two approaches were followed:
A top-free \ww{} cross section can be obtained in the four-flavour scheme (4FS)
by dropping all contributions with real bottom quarks, which are separately finite 
due to the bottom-quark mass.
Since in the five-flavour scheme (5FS)  
real and virtual contributions of massless bottom quarks are inevitably tied together,
the resonance structure of top-quark 
contributions is exploited 
to determine a top-free cross section.
Neither of the two approaches is required in the case of the \zz{} measurements presented here.
Since \ww{} and top-quark processes are both treated as backgrounds in the respective experimental analyses,
we introduce the following procedure: First, we compute the SF process 
$pp \rightarrow \ell^+\ell^-\,\nu_{\ell}\bar\nu_{\ell}$ including all resonant contributions.
In order to keep only \zz{} topologies (and interferences), we then subtract 
the DF process $pp \rightarrow \ell^+\nu_{\ell}\,\ell'^-\bar\nu_{\ell'}$. 
This removes \ww{} and top-quark backgrounds from our predictions, as desired,
while their interference with \zz{} production, which is not accounted for in the background predictions and thus considered part of the \zz{} signal, is kept. 
Its impact will be studied in detail below.
If \ww{} or top-quark topologies yield much larger contributions than \zz{} to the
SF process, sizeable cancellations in the subtraction could diminish the numerical 
accuracy of our predictions.
However, for typical \zz{} signal cuts, as considered here, a \z{}-mass 
window suppresses the \ww{} contribution, and a jet veto the top-quark background.
The presented procedure applies in all flavour schemes, and we conveniently use the 5FS 
throughout.

We present predictions for the 8\,TeV LHC. For the EW parameters we employ the 
$G_\mu$ scheme and compute the EW mixing angle as
$\cos\theta_W^2=(m_W^2-i\Gamma_W\,m_W)/(m_Z^2-i\Gamma_Z\,m_Z)$
and $\alpha=\sqrt{2}\,G_\mu m_W^2\sin^2\theta_W/\pi$,
using the complex-mass scheme~\cite{Denner:2005fg} throughout.
The EW inputs are set to the PDG~\cite{Patrignani:2016xqp} values: $G_F =
1.16639\times 10^{-5}$\,GeV$^{-2}$, $m_W=80.385$\,GeV,
$\Gamma_W=2.0854$\,GeV, $m_Z = 91.1876$\,GeV, $\Gamma_Z=2.4952$\,GeV,
$m_H = 125$\,GeV, and $\Gamma_H = 0.00407$.
The branching ratio of the $Z$-boson decay into massless charged leptons, $\ell\in\{e,\mu\}$, is 
$\textrm{BR}(Z \rightarrow \ell\ell) = 0.033631$, which is used 
below to compute the cross section in the total phase space.
The on-shell top-quark mass is set to $m_t = 173.2$\,GeV, and  
$\Gamma_t=1.44262$ is used.
For each perturbative order we use the corresponding set of $N_f=5$ 
NNPDF3.0~\cite{Ball:2014uwa} parton distributions
with $\as(m_Z)=0.118$.
Renormalization ($\mu_R$) and factorization ($\mu_F$) scales are set to
half of the invariant mass of the \zz{} pair, $\mu_R=\mu_F=\mu_0\equiv\frac{1}{2}\,m_{ZZ}$. Residual uncertainties are estimated from customary 7-point 
scale variations by a factor of two, with the constraint $0.5\le \mu_R/\mu_F\le 2$.

\renewcommand{\baselinestretch}{1.5}
\begin{table}[!b]
\begin{center}
\begin{tabular}{c}
\toprule
definition of the total phase space for $pp\to \zz{}+X$\\
\midrule
$66\,\textrm{GeV}\le m_{Z^{\rm rec}_{a/b}} \le 116$\,GeV\\
\bottomrule
definition of the fiducial volume for $pp\to \ell^+\ell^-\ell'^+\ell'^-+X,\quad \ell,\ell'\in\{e,\mu\}$\\
\midrule
$\ptl>7$\,GeV,  \quad one electron with $|\eta_e|<4.9$, \quad the others $|\eta_e|<2.5$,  \quad$|\eta_\mu|<2.7$\\
$\Delta R_{\ell\ell} >0.2$, \quad$\Delta R_{\ell\ell'} >0.2$, \quad$66\,\textrm{GeV}\le m_{Z^{\rm rec}_{a/b}} \le 116$\,GeV,\\
anti-$k_T$ jets with $R=0.4$, $p_{T,j}>25$\,GeV, $|\eta_j|<4.5$\\
lepton identification in SF channel:\\[-0.2cm]
minimizing differences of invariant-mass of OSSF lepton pairs and $m_Z$\\
\bottomrule
definition of the fiducial volume for $pp\to \ell^+\ell^-\nu\bar\nu+X,\quad \ell\in\{e,\mu\}$ and $\nu\in\{\nu_e,\nu_\mu,\nu_\tau\}$\\
\midrule
$\ptl>25$\,GeV, \quad$|\etal|<2.5$, \quad$\Delta R_{\ell\ell} >0.3$, \quad$76\,\textrm{GeV}\le \mll \le 106$\,GeV,\\
Axial-$\ptmiss>90$\,GeV, \quad$\pt\textrm{-balance}<0.4$,\\
$\njets=0$, \quad anti-$k_T$ %\cite{}
jets with $R=0.4$, $p_{T,j}>25$\,GeV, $|\eta_j|<4.5$ and $\Delta R_{ej}> 0.3$\\
\bottomrule
\end{tabular}
\end{center}
\renewcommand{\baselinestretch}{1.0}
\caption{\label{tab:cuts} %Definition of the fiducial volumes 
Phase-space definitions of the \zz{} measurements by ATLAS at 8\,TeV~\cite{Aaboud:2016urj}.}
\vspace{-0.5cm}
\end{table}

\renewcommand{\baselinestretch}{1.0}
We start by comparing phenomenological predictions to the ATLAS 8\,TeV measurement 
of \citere{Aaboud:2016urj}. The corresponding phase-space cuts are summarized in \tab{tab:cuts} for both the four-lepton and 
the $\ell\ell$+$E_T^{\rm miss}$ signatures.
The total phase space is defined by a $Z$-mass window in the invariant
mass of each reconstructed $Z$ boson. The reconstruction is unambiguous 
in the DF channel $\ell\ell\ell'\ell'$, $Z^{\rm rec}_a=\ell^+\ell^-$ and $Z^{\rm rec}_b=\ell'^+\ell'^-$,
which we employ for the predicted cross sections in the total phase space.
The fiducial cuts involve standard requirements 
on the transverse momenta and pseudo-rapidities of the leptons, a separation 
in $\Delta R=\sqrt{\Delta \eta^2+\Delta \phi^2}$ between the leptons, and 
a window in the invariant mass of reconstructed $Z$ bosons 
around the $Z$-pole. 
In the SF channel $\ell\ell\ell\ell$, $Z$ bosons are reconstructed by identifying the
combination of opposite-sign same-flavour (OSSF) lepton pairings ($Z_a=\ell_a^+\ell_a^-$ and $Z_b=\ell_b^+\ell_b^-$, or $Z_a=\ell_a^+\ell_b^-$ and $Z_b=\ell_b^+\ell_a^-$) that minimizes $\left| m_{Z_a}-m_Z\right|+\left|m_{Z_b}-m_Z\right|$ with the reconstructed $Z$ bosons $Z^{\rm rec}_a=Z_a$ and  $Z^{\rm rec}_b=Z_b$.
A rather special feature in the fiducial phase spaces of the 
four-lepton channels is the fact that ATLAS measures one of the electrons 
up to very large pseudo-rapidities ($|\eta_e|<4.9$). 
The measurement of the $\ell\ell$+$E_T^{\rm miss}$ signature applies two additional requirements, which 
force the two $Z$ bosons closer to back-to-back-like configurations to suppress backgrounds 
such as $Z$+jets: There is a lower cut on the axial missing transverse 
momentum,
$\text{Axial-}\ptmiss{} = -\ptmiss\cdot\cos\left(\Delta\phi_{\ell\ell,\nu\nu}\right)$,
where $\ptmiss \equiv p_{T,\nu\nu}$ and $\Delta\phi_{\ell\ell,\nu\nu}$ is the azimuthal angle 
between the dilepton and the neutrino pair. 
Furthermore, the two $Z$-boson momenta are balanced by putting 
an upper cut on $\pt{}\text{-balance}=|\ptmiss-\ptll|/\ptll$.
Finally, the $\ell\ell$+$E_T^{\rm miss}$ signature requires a jet veto to suppress top-quark backgrounds. Note that jets close to electrons ($\Delta R_{ej}<0.3$) are not vetoed.

In \tab{tab:ATLAS8} we report cross-section predictions and compare 
them against ATLAS 8\,TeV results~\cite{Aaboud:2016urj}.
Central predictions are stated with the numerical error on the last digit quoted in round 
brackets.
The relative uncertainties quoted in percent are estimated from 
scale variations as described above. Results reported for $e^+e^-\mu^+\mu^-$, 
$e^+e^-e^+e^-$, $\mu^+\mu^-\mu^+\mu^-$, $e^+e^-\nu\bar\nu$, and $\mu^+\mu^-\nu\bar\nu$ 
production are cross sections in the respective fiducial volumes defined in \tab{tab:cuts}.
The prediction in the last line of the table is obtained
from the computation of $pp\to e^+e^-\mu^+\mu^-+X$ in the total phase space
defined in \tab{tab:cuts}, 
by dividing out the branching ratio $\textrm{BR}(Z \rightarrow \ell\ell)$ for 
each $Z$-boson decay.
The main conclusions that can be drawn from these results are the following:

\renewcommand{\baselinestretch}{1.5}
\begin{table}[t]
\begin{center}
\resizebox{\columnwidth}{!}{%
\begin{tabular}{c c c c c}
\toprule
channel
& $\sigma_{\textrm{LO}}$ [fb]
& $\sigma_{\textrm{NLO}}$ [fb]
& $\sigma_{\textrm{NNLO}}$ [fb]
& $\sigma_{\textrm{ATLAS}}$ [fb]\\

\midrule
$e^+e^-\mu^+\mu^-$ & $8.188(1)_{-3.2\%}^{+2.4\%}$ & $11.30(0)_{-2.0\%}^{+2.5\%}$ & $12.92(1)_{-2.2\%}^{+2.8\%}$ 
& $12.4\;^{+1.0}_{-1.0}{\rm (stat)}\;^{+0.6}_{-0.5}{\rm (syst)}\;^{+0.3}_{-0.2}{\rm (lumi)}$\Bstrut\\
$e^+e^-e^+e^-$ & $4.654(0)_{-3.1\%}^{+2.3\%}$ & $6.410(2)_{-2.0\%}^{+2.5\%}$ & $7.310(8)_{-2.1\%}^{+2.7\%}$ 
& $5.9\;^{+0.8}_{-0.8}{\rm (stat)}\;^{+0.4}_{-0.4}{\rm (syst)}\pm 0.1{\rm (lumi)}$\Bstrut\\
$\mu^+\mu^-\mu^+\mu^-$ & $3.565(0)_{-3.5\%}^{+2.6\%}$ & $4.969(5)_{-2.0\%}^{+2.5\%}$ & $5.688(6)_{-2.2\%}^{+2.9\%}$ 
& $4.9\;^{+0.6}_{-0.5}{\rm (stat)}\;^{+0.3}_{-0.2}{\rm (syst)}\pm 0.1{\rm (lumi)}$\Bstrut\\
$e^+e^-\nu\nu$ & $5.558(0)_{-0.5\%}^{+0.1\%}$ & $4.806(1)_{-3.9\%}^{+3.5\%}$ & $5.083(8)_{-0.6\%}^{+1.9\%}$ 
& $5.0\;^{+0.8}_{-0.7}{\rm (stat)}\;^{+0.5}_{-0.4}{\rm (syst)}\pm 0.1{\rm (lumi)}$\Bstrut\\
$\mu^+\mu^-\nu\nu$ & $5.558(0)_{-0.5\%}^{+0.1\%}$ & $4.770(4)_{-4.0\%}^{+3.6\%}$ & $5.035(9)_{-0.5\%}^{+1.8\%}$ 
& $4.7\;^{+0.7}_{-0.7}{\rm (stat)}\;^{+0.5}_{-0.4}{\rm (syst)}\pm 0.1{\rm (lumi)}$\Bstrut\\
\midrule
total rate & $4982(0)_{-2.7\%}^{+1.9\%}$ & $6754(2)_{-2.0\%}^{+2.4\%}$ & $7690(5)_{-2.1\%}^{+2.7\%}$ 
& $7300\;^{+400}_{-400}{\rm (stat)}\;^{+300}_{-300}{\rm (syst)}\;^{+200}_{-100}{\rm (lumi)}$\Bstrut\\
\bottomrule

\end{tabular}}
\end{center}
\renewcommand{\baselinestretch}{1.0}
\caption{\label{tab:ATLAS8} Predictions for fiducial and total rates compared to ATLAS 8 TeV data~\cite{Aaboud:2016urj}.}
\end{table}

\renewcommand{\baselinestretch}{1.0}
\vspace{-0.35cm}
\begin{itemize}
 \itemsep0em
\item Radiative corrections are large and have a marked dependence on the
event selection: They range between $+35$\% to $+40$\% at NLO and
$+14$\% to $+17$\% at NNLO in cases without a jet veto, i.e.\ for all but the $2\ell2\nu$ results. 
Roughly half ($45$\%--$55$\%) of the $\mathcal{O}(\alpha_s^2)$ terms are due to the 
loop-induced $gg$ component in these cases. 
For the $2\ell2\nu$ processes 
the situation is quite different: Due to the jet veto
NLO corrections turn negative and yield about $-14\%$. 
NNLO corrections are roughly $+6\%$. However, the positive effect is 
entirely due to loop-induced $gg$ contributions, which are not affected
by the jet veto. Omitting the loop-induced $gg$ terms, the genuine NNLO corrections 
to the $q\bar{q}$ channel are actually negative and about $-5\%$.
Hence, despite the jet veto, full $\mathcal{O}(\alpha_s^2)$ corrections are crucial 
for the $\ell\ell$+$E_T^{\rm miss}$ signature.
\item For channels with four charged leptons we find good agreement between 
theory and data. This is particularly true for the DF process ($e^+e^-\,\mu^+\mu^-$),
where NNLO corrections clearly improve the comparison. In the SF 
channels ($e^+e^-\,e^+e^-$ and $\mu^+\mu^-\,\mu^+\mu^-$) NNLO predictions 
are slightly larger than the measurements, but remain within $1\sigma$ for muons 
and $2\sigma$ for electrons. One should not forget that EW corrections 
reduce the rates by a few percent~\cite{Biedermann:2016lvg}, while NLO corrections to the
loop-induced $gg$ channel have a positive effect~\cite{Caola:2015psa}.
\item For the $\ell\ell$+$E_T^{\rm miss}$ signatures excellent agreement is found
between NNLO predictions and measured cross sections. It is worth noting 
that fixed-order results describe the data significantly better than the \powheg{} \cite{Nason:2004rx,Frixione:2007vw,Alioli:2010xd,Melia:2011tj} 
Monte Carlo prediction used in \citere{Aaboud:2016urj}. This could be 
caused by the jet-veto requirement: As pointed out in \citere{Monni:2014zra} 
for \ww{} production, in presence of a jet veto the fiducial rate predicted by \powheg{} is rather small.
\item The NNLO prediction in the last line of the table
agrees perfectly ($<1\sigma$) with the experimental result in the total phase space, 
with NNLO corrections being crucial for this level of agreement.
\item At LO scale uncertainties  
clearly underestimate the actual size of higher-order corrections, 
since only the $q\bar q$ channel contributes and the cross section is $\mu_R$-independent. 
Given large NLO corrections, also the 
scale uncertainties of $2$\%--$4$\% at NLO cannot be trusted as an
estimate of missing higher-order terms.
However, at NNLO all partonic channels are included, and
the corrections to the $q\bar q$ channel, which are much smaller than at NLO,
are of the same order as the respective scale variations. 
Therefore, NNLO uncertainties may be expected to reflect the 
size of yet un-calculated perturbative corrections to this channel. 
Only the loop-induced $gg$ component underestimates the uncertainty due to its LO 
nature, which is known from the sizable NLO contributions to the $gg$ channel \cite{Caola:2015psa}.
\end{itemize}

\begin{figure}
\begin{center}
\begin{tabular}{cc}
\includegraphics[trim = 7mm -7mm 0mm 0mm, width=.33\textheight]{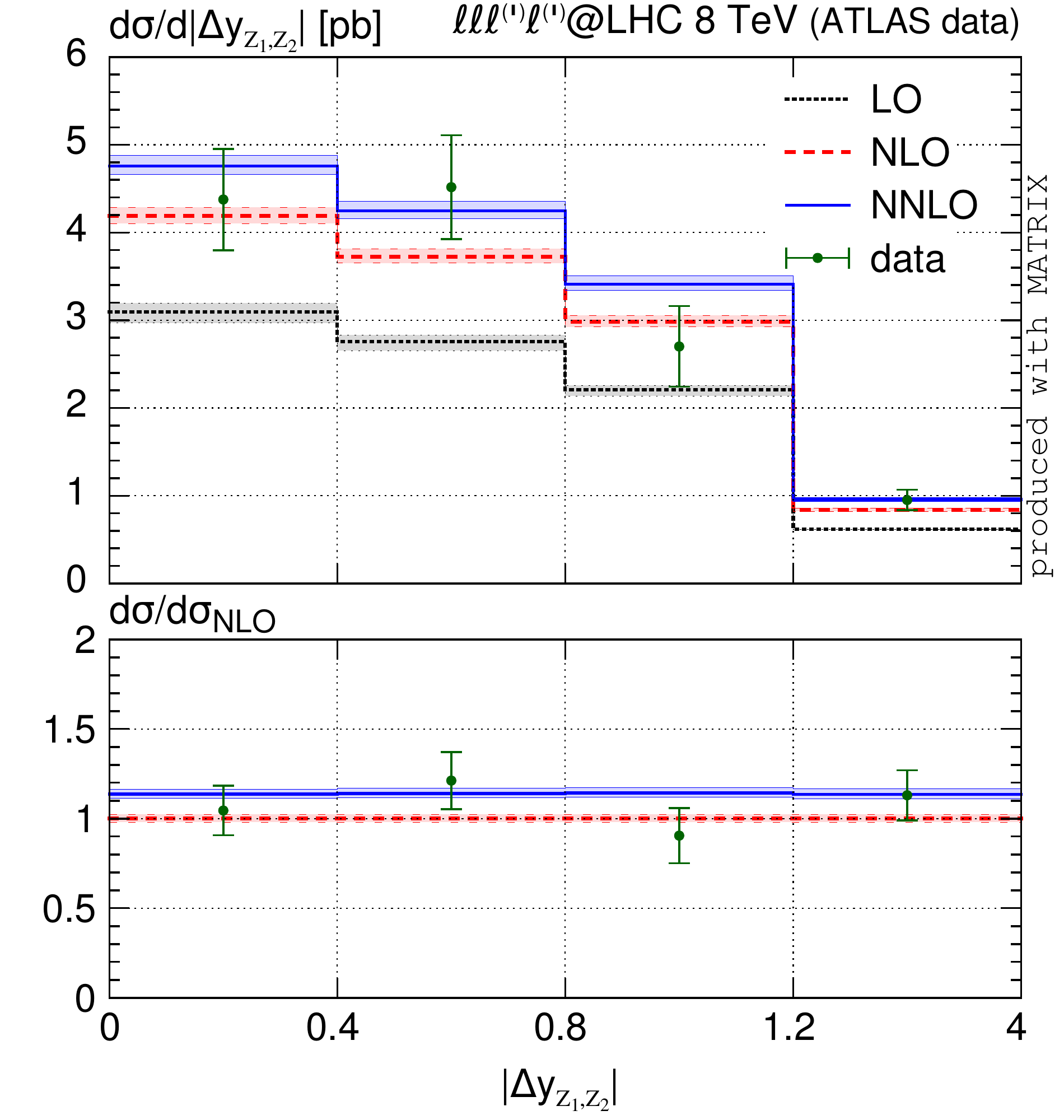} &
\includegraphics[trim = 7mm -7mm 0mm 0mm, width=.33\textheight]{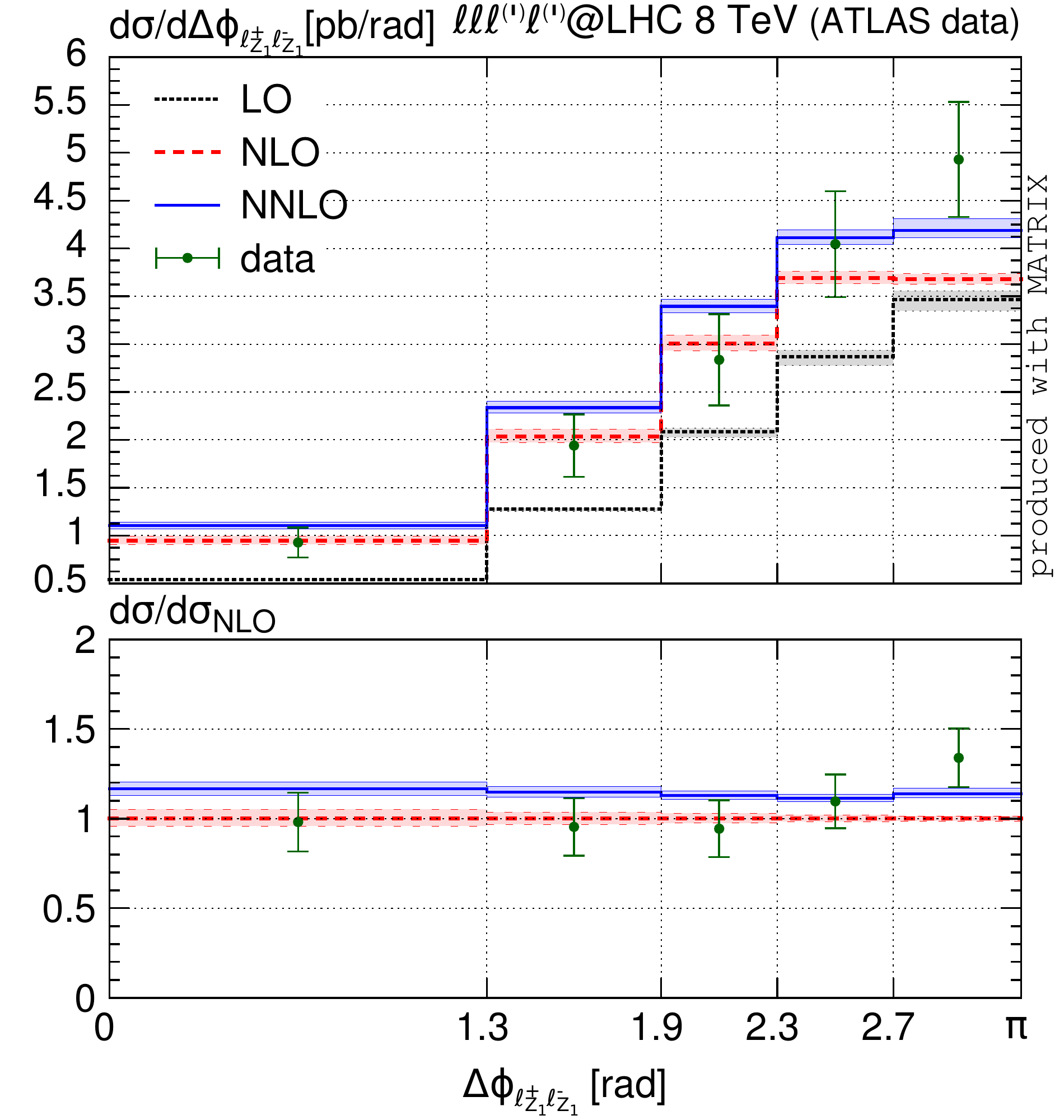} \\[-0.8em]
\hspace{0.6em} (a) & \hspace{1em}(b)
\end{tabular}\vspace{0.5cm}
\begin{tabular}{cc}
\includegraphics[trim = 7mm -7mm 0mm 0mm, width=.33\textheight]{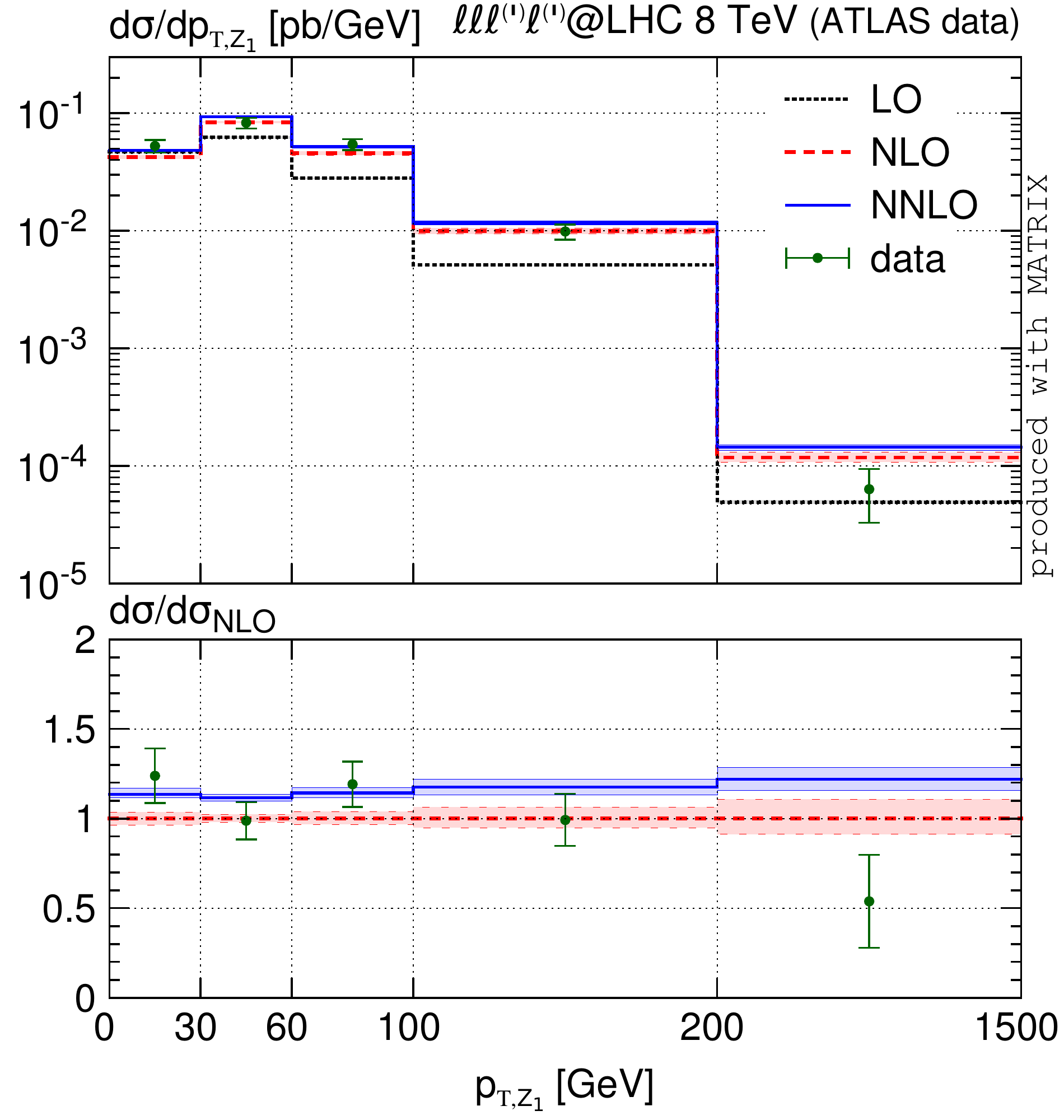} &
\includegraphics[trim = 7mm -7mm 0mm 0mm, width=.33\textheight]{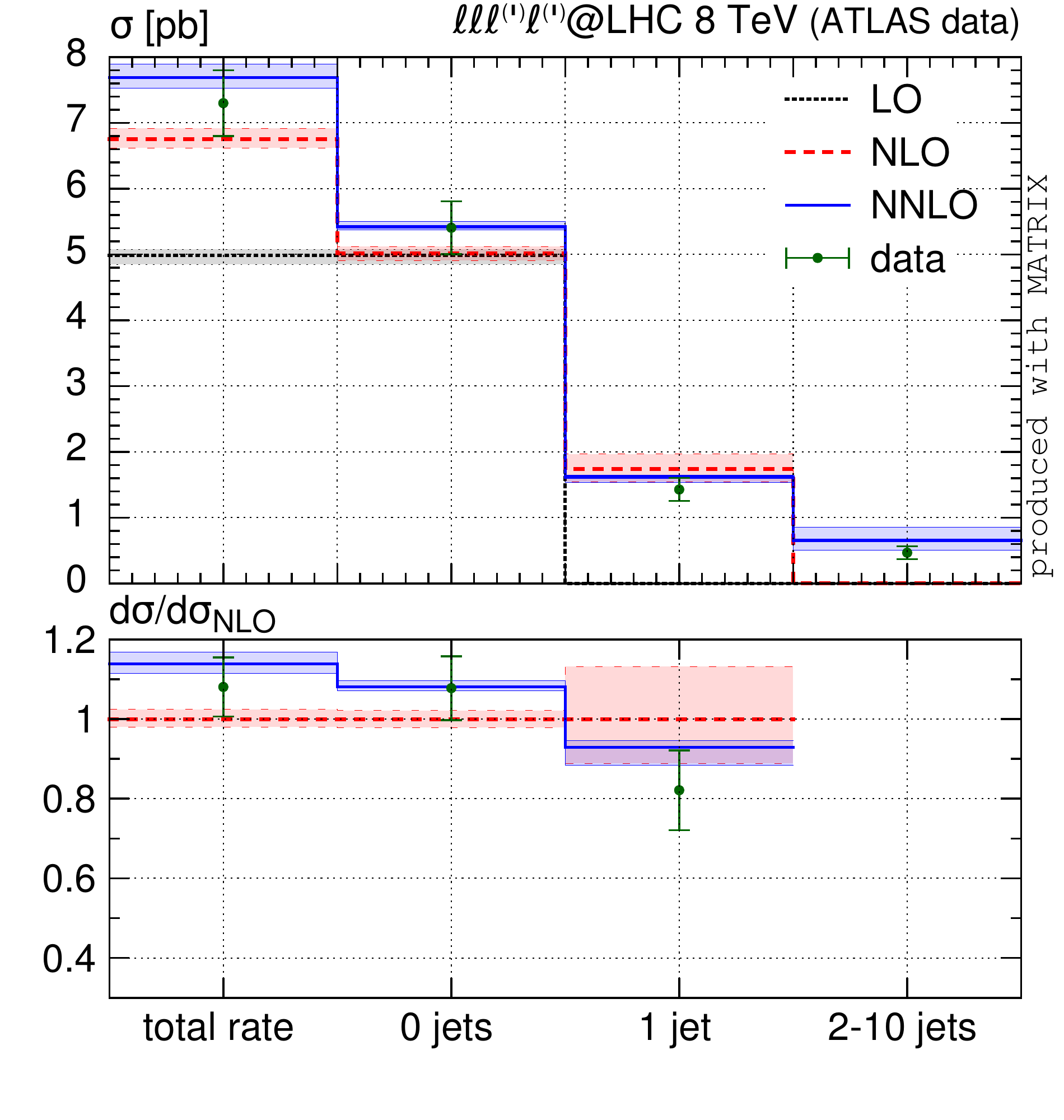} \\[-0.8em]
\hspace{0.6em} (c) & \hspace{1em}(d)
\end{tabular}\vspace{0.5cm}
\caption[]{{
\label{tab:4l}
Differential distributions for the four-lepton processes in the total phase space at LO (black, dotted), 
NLO (red, dashed) and NNLO (blue, solid), compared to 
ATLAS 8\,TeV data extrapolated to the total phase space \cite{Aaboud:2016urj} (green points with error bars); 
for (a) \dyzz{}, (b) \dphilzlz{}, 
(c) \ptzone{}, and (d) \njets{}; the lower frames show the 
ratio over NLO.}}
\end{center}
\end{figure}

We now turn to discussing differential distributions. \fig{tab:4l} shows results for the production of 
four charged leptons in the total phase space. Theoretical predictions in these plots are obtained
from the DF process $pp\to e^+e^-\,\mu^+\mu^-+X$, divided by the branching 
ratio $\textrm{BR}(Z \rightarrow \ell\ell)$ for each $Z$-boson decay.
The measured results are extrapolated to the total phase space, as presented by ATLAS at 8\,TeV~\cite{Aaboud:2016urj}.
Given that one electron is measured up to absolute pseudo-rapidities of $4.9$, the extrapolation factor, 
and possibly the ensuing uncertainty, is smaller than in other four-lepton measurements. 
Nevertheless, we reckon that a direct comparison against unfolded distributions 
in the fiducial 
volume is preferable, as it is less affected by the lower perturbative accuracy of the Monte Carlo generator used for 
the extrapolation. However, since no such experimental results are available in the four-lepton channel 
from ATLAS at 8\,TeV, we perform the comparison in the total phase space. We have normalized the 
ATLAS distributions to the measured total cross section in the last line of \tab{tab:ATLAS8}.

Despite the fact that the comparison is done in the total phase space, theory predictions and measured cross sections are in 
reasonable agreement for the observables shown in \fig{tab:4l}, which are 
the rapidity difference of the reconstructed $Z$ bosons, \dyzz{}~(panel a), the azimuthal angle 
between the two leptons of the harder $Z$ boson, \dphilzlz{}~(panel b), the transverse momentum of the leading 
$Z$ boson, \ptzone{}~(panel c), and the number of jets, \njets{}~(panel d). 
Overall, NNLO predictions provide the best description of data, although NLO results are
similarly close, while LO is far off.
Note that for the jet multiplicity the effective perturbative accuracy of the (fixed-order) predictions 
is degraded by one order for each added jet.
NNLO effects on other distributions 
are large, but primarily affect the normalization and not the shapes.

\begin{figure}[t]
\begin{center}
\begin{tabular}{ccc}
\hspace{-0.03cm}\includegraphics[trim = 7mm -7mm 0mm 0mm, width=.24\textheight]{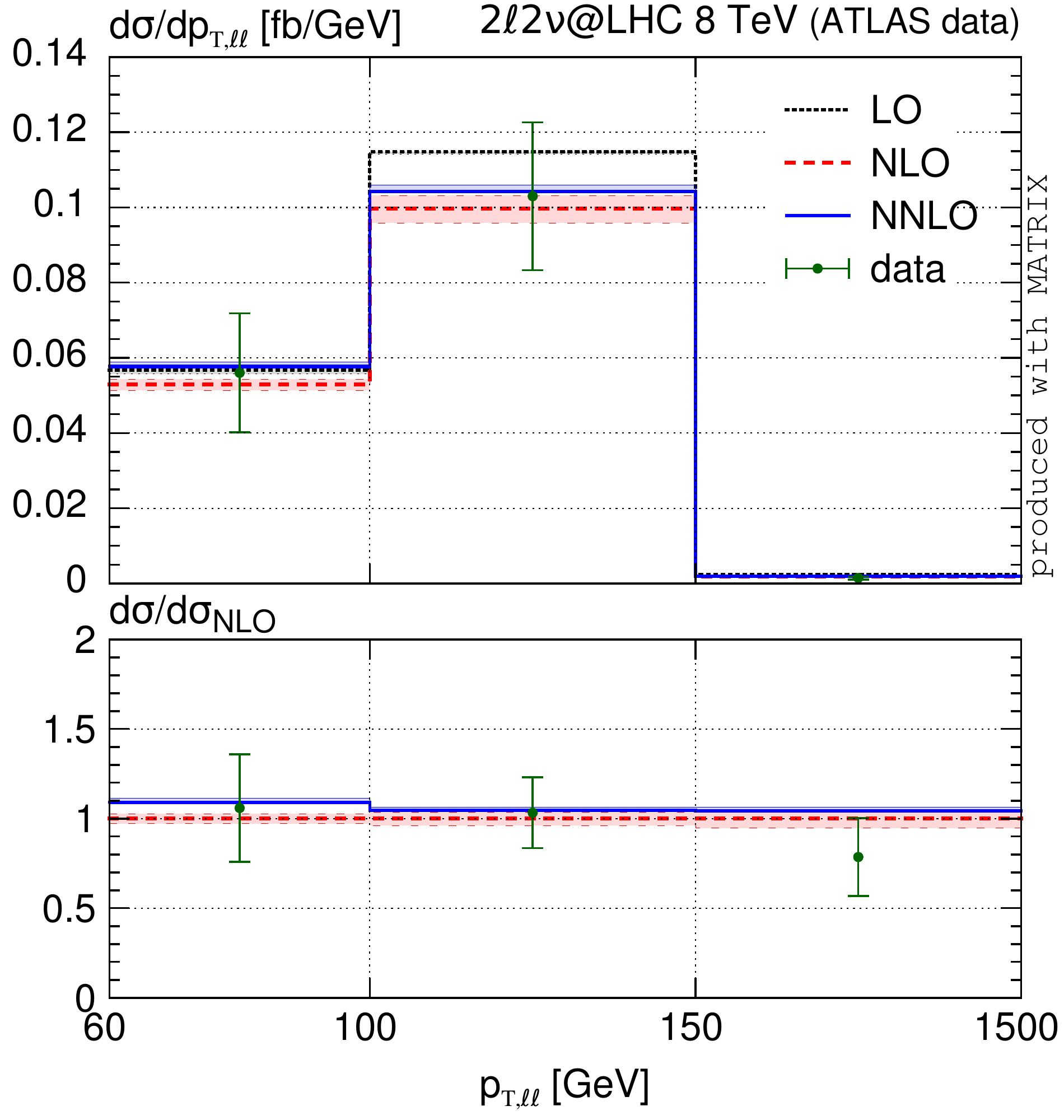} &
\hspace{-0.1cm}\includegraphics[trim = 7mm -7mm 0mm 0mm, width=.24\textheight]{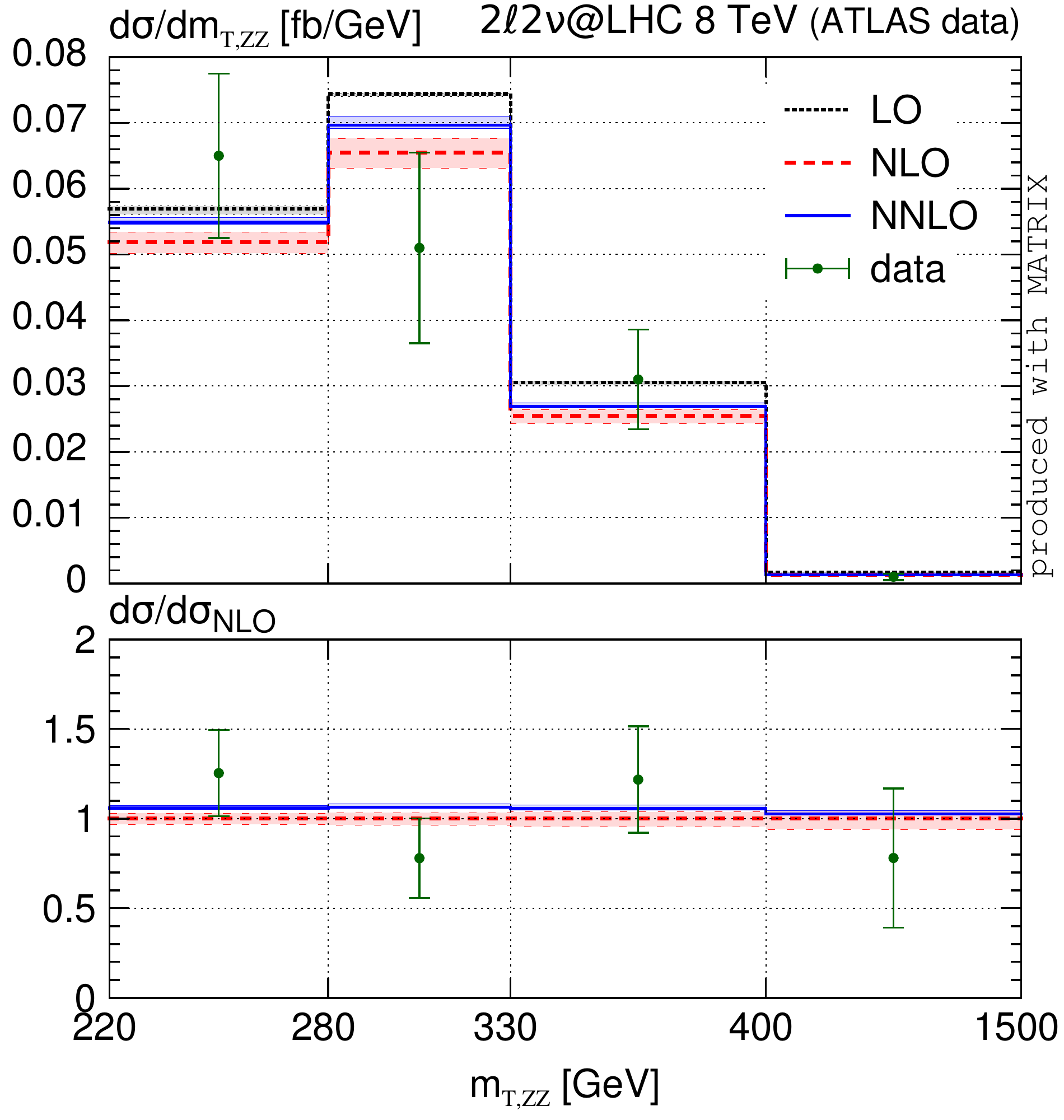} &
\hspace{-0.3cm}\includegraphics[trim = 7mm -7mm 0mm 0mm, width=.24\textheight]{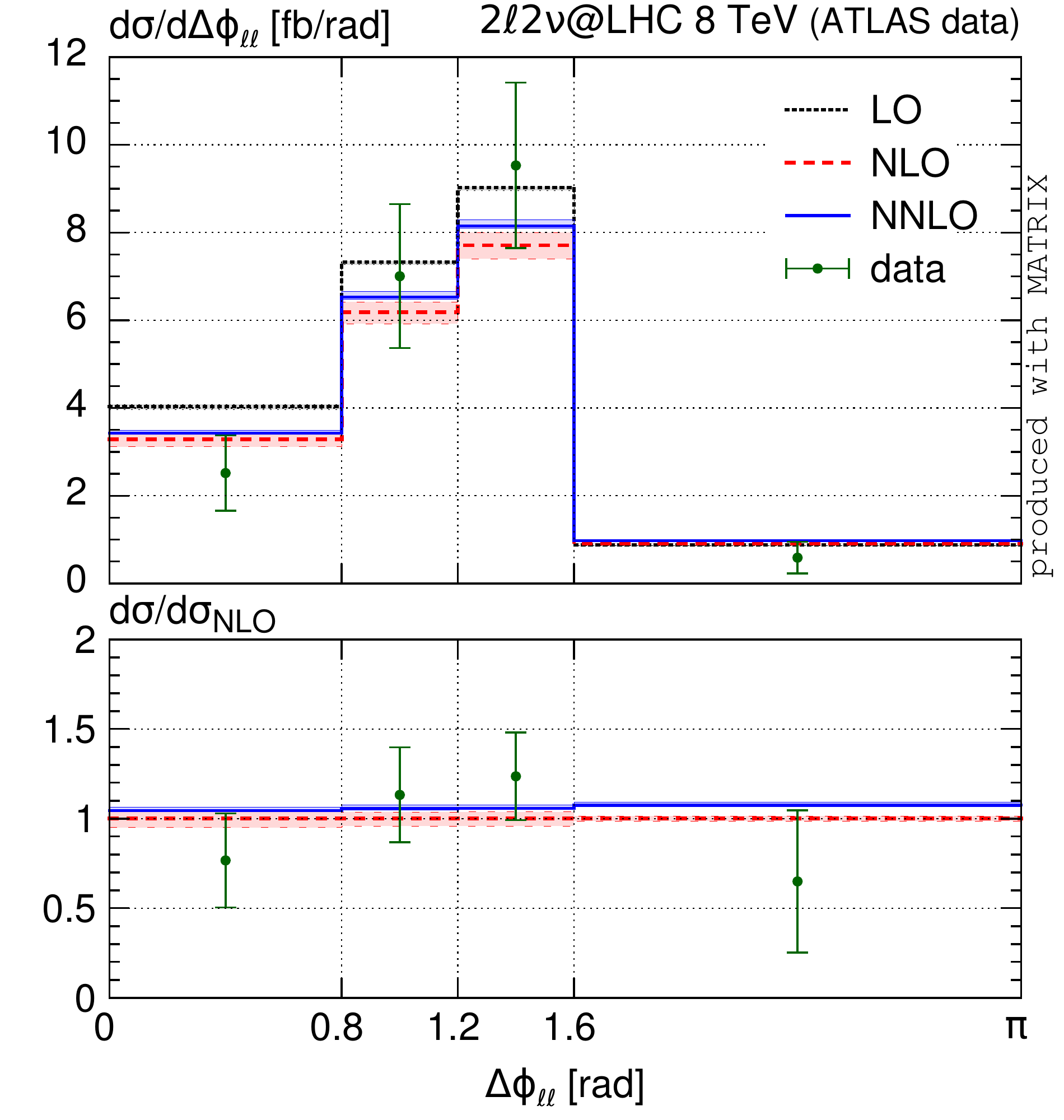} \\[-0.8em]
\hspace{0.1em} (a) & \hspace{0.2em}(b) & \hspace{0.2em}(c)
\end{tabular}\vspace{0.5cm}
\caption[]{\label{fig:llnunu}{
Differential distributions of the $2\ell2\nu$ processes with fiducial cuts at LO (black, dotted), 
NLO (red, dashed) and NNLO (blue, solid), compared to 
ATLAS 8\,TeV data  \cite{Aaboud:2016urj} (green points with error bars); 
for (a) \ptll{}, (b) \mtzz{}, and (c) \dphill{}; the lower frame shows the 
ratio over NLO.}}
\end{center}
\end{figure}

We continue our discussion of differential results 
with the $\ell\ell$+$E_T^{\rm miss}$ signature in \fig{fig:llnunu}, which shows the 
distributions in the transverse momentum of the dilepton pair, \ptll{}~(panel a), the transverse mass of the $ZZ$ 
pair, defined as\footnote{Boldface is 
used to indicate the vectorial sum of the dilepton and missing transverse momentum.}
\begin{align}
\label{eq:mTW}
\mtzz = \sqrt{\left(\sqrt{\ptll^2+m_Z^2} + \sqrt{(\ptmiss)^2+m_Z^2}\right)^2 - ({{\bf p}_{{\bf T},\ell\ell}+{\bf p}_{\bf T}^{\rm miss}})^2}\nonumber
\end{align}
(panel b), and the azimuthal angle between the two leptons, 
\dphill{}~(panel c). The results correspond to 
the sum of all channels 
including both SF ($\ell\ell\,\nu_{\ell}\nu_{\ell}$) and DF ($\ell\ell\,\nu_{\ell'}\nu_{\ell'}$) processes
($\ell\in\{e,\mu\},\,\nu_{\ell'}\in\{\nu_e,\nu_\mu,\nu_\tau\},\,\ell\neq\ell'$). 
We recall that SF contributions are computed by subtracting \ww{} and top-quark 
backgrounds as outlined before. For all three distributions in 
\fig{fig:llnunu} we find excellent agreement between theory and data.
At NNLO, differences hardly exceed the $1\sigma$ level.
Although NNLO corrections change the cross section in certain bins, the experimental uncertainties are still too large
for more distinct conclusions. Similar to our previous observations for fiducial rates, the agreement found here at 
fixed order is a significant improvement over the comparison with the Monte Carlo prediction shown 
in \citere{Aaboud:2016urj}. As pointed out before,
we expect a poor modelling of the jet veto by the \powheg{} generator to be the main source of these differences, see also \citere{Monni:2014zra}.

\begin{figure}[t]
\begin{center}
\begin{tabular}{ccc}
\hspace{-0.1cm}\includegraphics[trim = 7mm -7mm 0mm 0mm, width=.24\textheight]{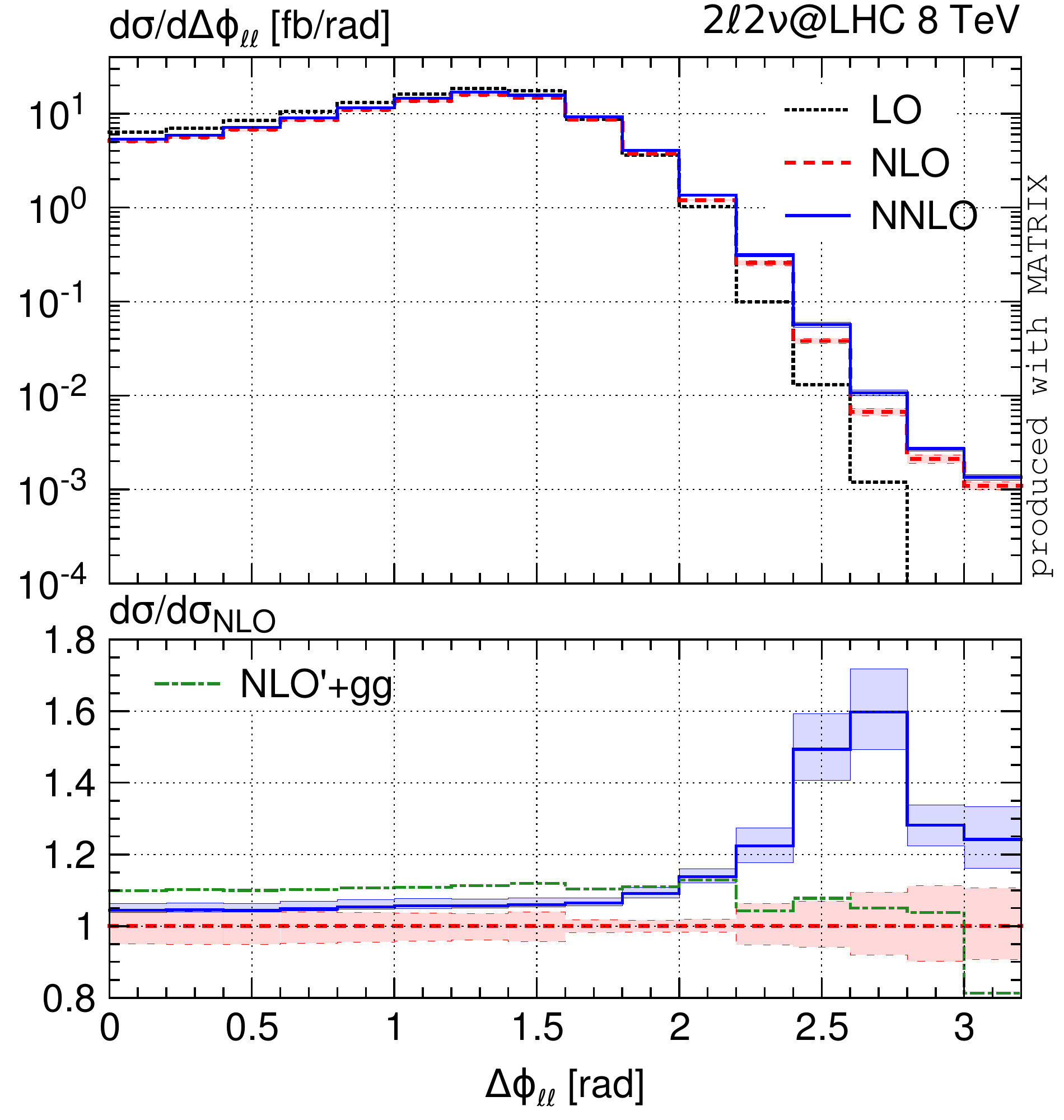} &
\hspace{-0.25cm}\includegraphics[trim = 7mm -7mm 0mm 0mm, width=.24\textheight]{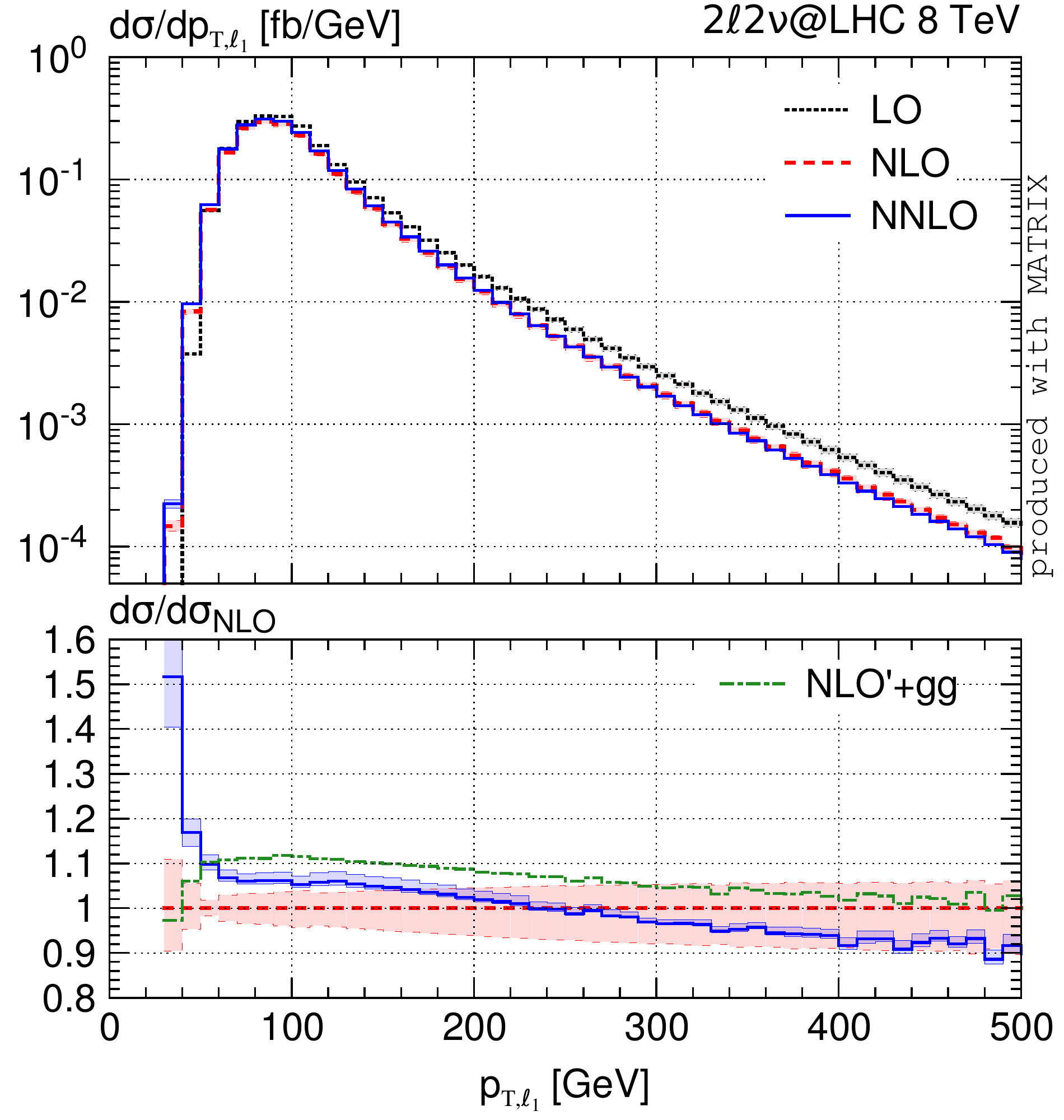} &
\hspace{-0.25cm}\includegraphics[trim = 7mm -7mm 0mm 0mm, width=.24\textheight]{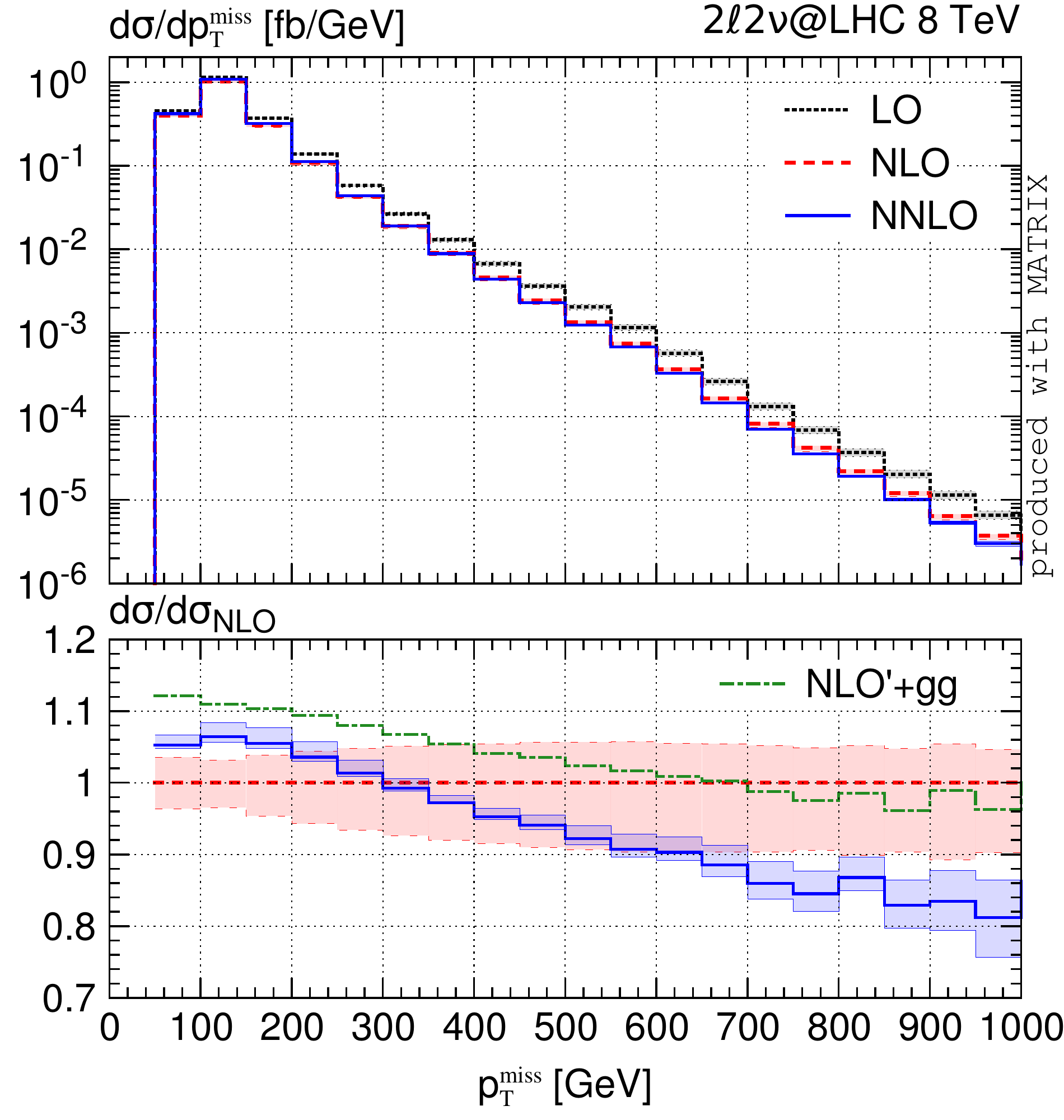} \\[-0.8em]
\hspace{0.1em} (a) & \hspace{0.2em}(b) & \hspace{0.2em}(c)
\end{tabular}\vspace{0.5cm}
\caption[]{\label{fig:2l2nuother}{Same as \fig{fig:llnunu}, but without data and for the distributions (a) \dphill{}, (b) \ptlone{}, and (c) \ptmiss{}; for
reference, also the NLO$^\prime$+$gg$ result (green, dash-dotted) is shown in the ratio frame.}}
\end{center}
\end{figure}

In the remainder of this paper we focus on the $\ell\ell$+$E_T^{\rm miss}$ 
signature, with the same fiducial setup as before. 
In \fig{fig:2l2nuother} we have picked three out of many observables where 
the importance of NNLO corrections is evident. 
The NLO$^\prime$+$gg$ result in the ratio frame denotes the sum of the 
NLO and the loop-induced $gg$ cross section, both evaluated with NNLO PDFs,
which was the best prediction available in the past. 
Its difference compared to the complete NNLO QCD result shows the size of the genuine $\mathcal{O}(\as^2)$
corrections to the $q\bar{q}$ channel, computed for the first time in this paper.
For example, the \dphill{} distribution in \fig{fig:2l2nuother}~(panel a)
develops a sizable NNLO/NLO $K$-factor up to $1.6$ for large separations. 
From the considerable differences between NNLO and NLO$^\prime$+$gg$ curves,
which also concern their shapes, it is clear that this effect stems directly from the newly 
computed $\mathcal{O}(\as^2)$ contributions.
In this phase-space region (large \dphill{}) the perturbative accuracy is effectively 
diminished by one order due to the phase-space cuts which force the two $Z$ bosons to be
boosted and approximately back-to-back, so that the two decay leptons disfavour large separations.
This manifests itself also in a widening of the scale uncertainty bands. Also the 
transverse-momentum spectrum of the hardest lepton, \ptlone{} in \fig{fig:2l2nuother}~(panel b)
features a significant shape distortion at NNLO, when compared to both NLO and 
NLO$^\prime$+$gg$. The same is true for the missing transverse momentum, \ptmiss{}
in \fig{fig:2l2nuother}~(panel c). In all cases perturbative uncertainties are clearly reduced 
upon inclusion of higher-order corrections.

\begin{figure}[t]
\begin{center}
\begin{tabular}{ccc}
\includegraphics[trim = 7mm -7mm 0mm 0mm, width=.24\textheight]{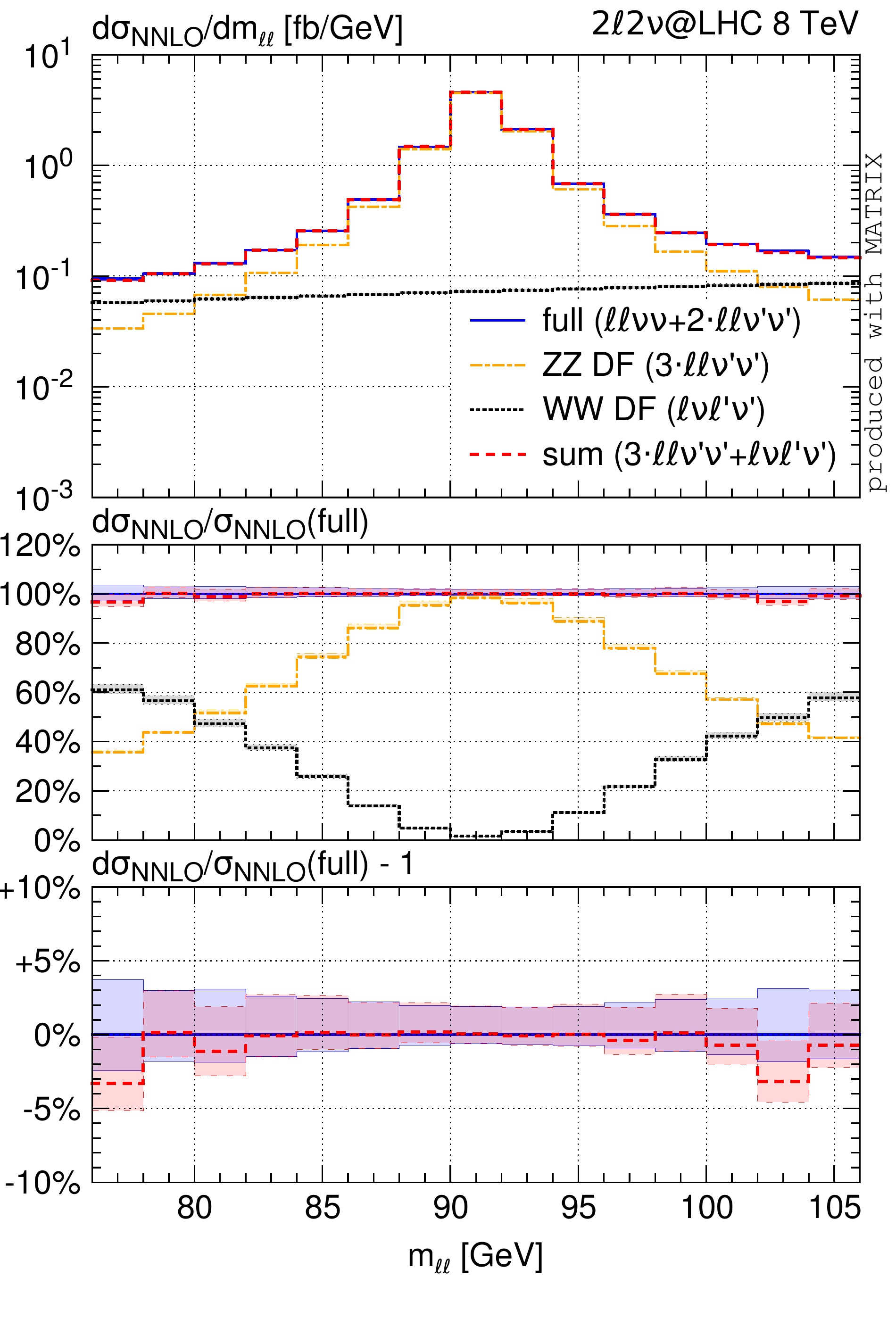} &
\hspace{-0.25cm}\includegraphics[trim = 7mm -7mm 0mm 0mm, width=.24\textheight]{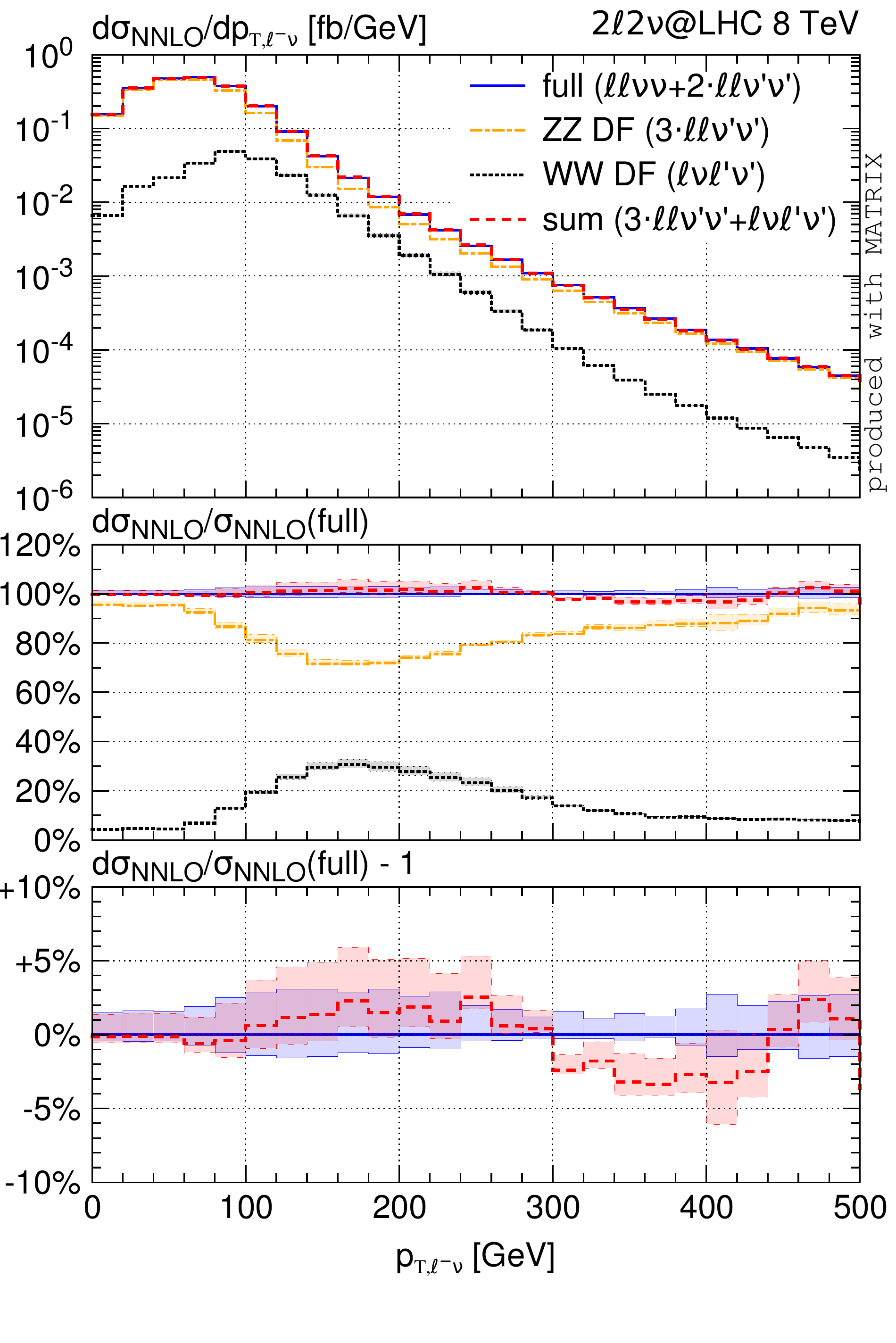} &
\hspace{-0.25cm}\includegraphics[trim = 7mm -7mm 0mm 0mm, width=.24\textheight]{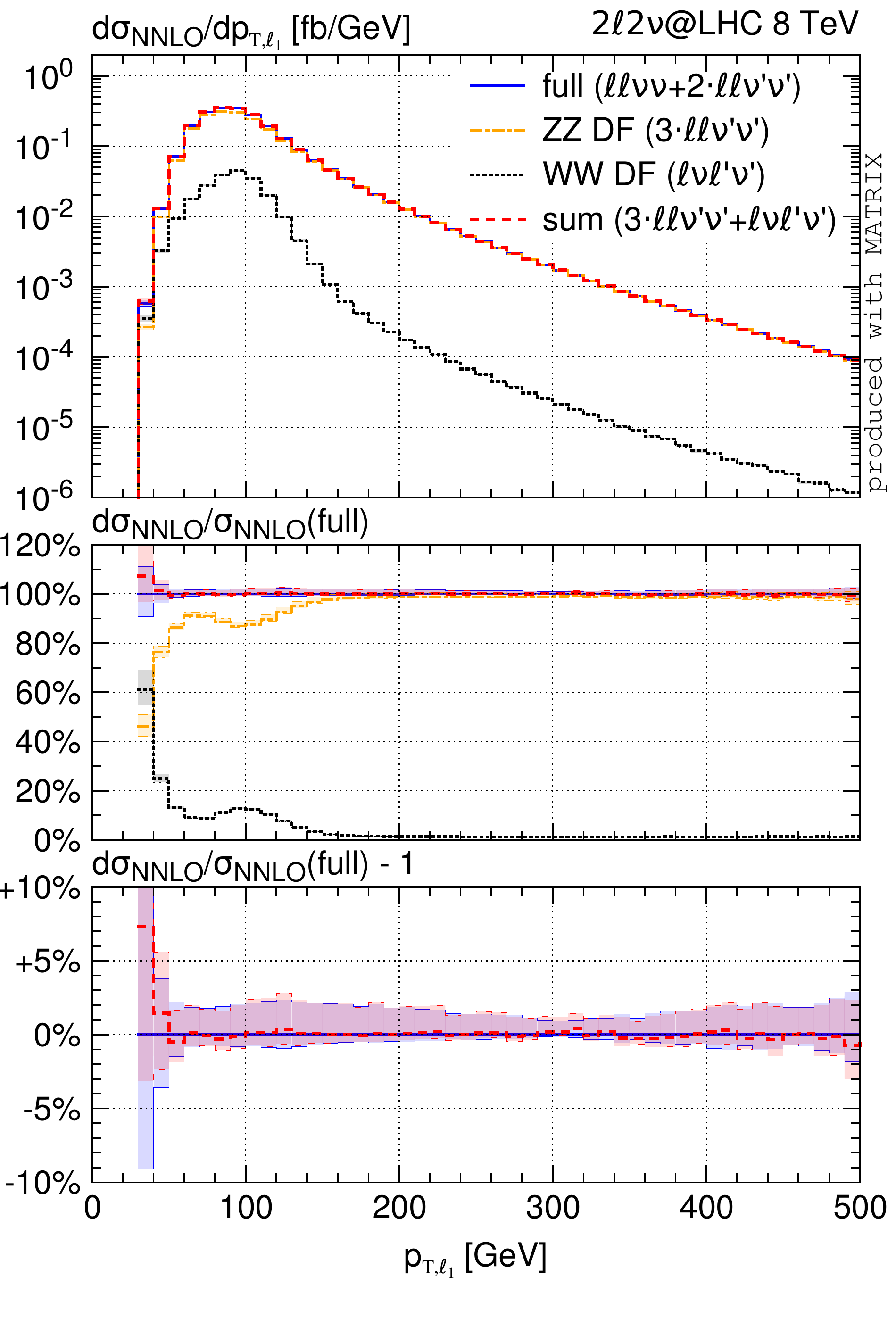} \\[-1.7em]
\hspace{0.1em} (a) & \hspace{0.2em}(b) & \hspace{0.2em}(c)
\end{tabular}\vspace{0.5cm}
\caption[]{\label{fig:ZZvsWW}{
Comparison of NNLO cross sections for the full process 
$\sigma(\ell\ell\,\nu_{e/\mu/\tau}\nu_{e/\mu/\tau})$ (blue, solid), the individual \zz{} contributions $3\cdot\sigma(\ell\ell\,\nu_{\ell'}\nu_{\ell'})$ with $\ell\neq\ell$ (orange, dash-dotted), 
the individual \ww{} contributions $\sigma(\ell\nu_\ell\,\ell'\nu_{\ell'})$ with $\ell\neq\ell$ (black, dotted), and the approximation of the full result by the 
incoherent sum of \zz{} and \ww{} contributions  $3\cdot\sigma(\ell\ell\,\nu_{\ell'}\nu_{\ell'})+\sigma(\ell\ell'\,\nu_{\ell}\nu_{\ell'})$ (red, dashed); for (a) \mll{}, (b) \ptwm{}, and (c) \ptlone{}; the lower frames show the ratio to the 
full result.}}
\end{center}
\end{figure}

We complete our discussion of phenomenological results by studying 
the size of \zz{}, \ww{}, and interference contributions entering 
the SF process $pp \rightarrow \ell^+\ell^-\,\nu_{\ell}\bar\nu_{\ell}$. We recall that
\ww{} contributions also involve resonant top-quark topologies. 
In contrast to our previous discussion, \ww{} and top-quark 
contributions are not subtracted from the SF process in the following. 
We focus on the contamination of the $\ell\ell$+$E_T^{\rm miss}$ signature through 
interference with \ww{} and top-quark diagrams.
To this end, \fig{fig:ZZvsWW} compares the 
NNLO cross section 
for the full process of two OSSF leptons and two neutrinos,
$\sigma({\ell\ell\,\nu_{e/\mu/\tau}\nu_{e/\mu/\tau}})=\sigma({\ell\ell\,\nu_{\ell}\nu_{\ell}})+2\cdot \sigma({\ell\ell\,\nu_{\ell'}\nu_{\ell'}})$ for $\ell\in\{e,\mu\}$ and $\ell\neq\ell'$ with the 
same NNLO cross section, where the SF channel is approximated by the incoherent sum 
of the two DF processes, $\sigma({\ell\ell\,\nu_{e/\mu/\tau}\nu_{e/\mu/\tau}})\approx3\cdot \sigma({\ell\ell\,\nu_{\ell'}\nu_{\ell'}})+\sigma({\ell\nu_{\ell}\,\ell'\nu_{\ell'}})$. 
The difference of the two is precisely the remaining 
interference contribution of \zz{} with \ww{} (and top-quark) topologies which we 
want to study. For completeness, also the individual DF \zz{} and DF \ww{} cross sections, 
$3\cdot \sigma({\ell\ell\,\nu_{\ell'}\nu_{\ell'}})$ and $\sigma({\ell\ell'\,\nu_{\ell}\nu_{\ell'}})$, respectively, are shown, whose sum is the approximated cross section.

It is instructive to consider
the invariant mass of the charged leptons, \mll{}, in \fig{fig:ZZvsWW}~(panel a),
which nicely illustrates the nature of the different results: 
Only \zz{} topologies feature a resonance at 
$\mll{}=\mz{}$, while the DF \ww{} prediction is almost flat in this range of \mll{}.
It is clear from the first ratio frame that almost the entire cross section 
around the peak stems from \zz{} contributions. Only away from the peak 
\ww{} production becomes larger than \zz{} production. It is also clear
that it is the \mll{} cut in the fiducial definition which significantly enhances \zz{}
contributions and suppresses the \ww{} process. The relative difference between 
the approximated and the full 
result, which is enlarged in the second ratio frame, is very small, in particular 
in the peak region. This demonstrates that interference effects of \zz{} with 
\ww{} (and top-quark) topologies are negligible, and that an incoherent sum of 
the two DF channels is an excellent approximation of the SF process.
This also implies that in our previous definition of the $\ell\ell$+$E_T^{\rm miss}$ signature
the remaining interference effects after subtraction of \ww{} and top-quark backgrounds are small.
In fact, we hardly found any distribution with larger interference effects. 
The most pronounced example is the ``pseudo''-observable 
in \fig{fig:ZZvsWW}~(panel b) that shows the transverse-momentum 
spectrum of a $W^-$ boson reconstructed as $\ell^-\nu_\ell$, and even in this 
case the differences do not exceed a few percent, although the shape is slightly 
deformed. With interference effects being generally small, it is interesting 
to analyse the different behaviour of \zz{} and \ww{} topologies.
In the \ptlone{} distribution in \fig{fig:ZZvsWW}~(panel c), for example, the 
relative \ww{} contribution increases around $\ptlone=90$\,GeV. 
This feature is already present at LO, and it is caused by purely kinematic effects that 
allow the two $W$ bosons to become resonant simultaneously only in this part of phase space.
The region below $\ptlone=45$\,GeV is populated only beyond LO.

We have presented \nnlo{} QCD corrections to \zz{} production for all
leptonic processes. The $\ell\ell$+$E_T^{\rm miss}$ signature has been studied for the 
first time at this level of accuracy, and we have introduced a procedure to compute results consistently
in the five-flavour scheme without contributions from \ww{} or top-quark backgrounds.
We also computed state-of-the-art predictions for signatures involving four charged leptons.
Our results are compared to ATLAS data at 8\,TeV, and we find good agreement 
for both fiducial cross sections and distributions. 
NNLO QCD corrections are sizable, 
even in presence of a jet veto used 
in the $\ell\ell$+$E_T^{\rm miss}$ measurement. 
By and large, they are of the order of $5$\%--$20$\%, but can reach even 
$60\%$ in certain phase-space regions. Most importantly, such effects do not only stem 
from the loop-induced $gg$ contribution, but are also due to the newly computed
genuine $\mathcal{O}(\as^2)$ corrections to the $q\bar{q}$ channel.
Not least, we have shown that all remaining interference
effects of \zz{} topologies with \ww{} and top-quark backgrounds in $2\ell 2\nu$ production are 
negligible. The availability of fully differential NNLO predictions for all leptonic channels of 
\zz{} production will play a crucial role 
in the rich physics programme that is based on precision studies of \zz{} signatures at the LHC.
Along with the paper we provide an updated version of \Matrix{},
featuring all processes with the fiducial setup, cuts and distributions considered here.

\noindent {\bf Acknowledgements.}
We would like to thank Massimiliano Grazzini and Jochen Meyer
 for useful discussions and comments on the manuscript.
The work of MW is supported by the ERC Consolidator Grant 614577 HICCUP.
\vspace{0.1cm}
\setlength{\bibsep}{3.5pt}
\renewcommand{\em}{}
\bibliographystyle{apsrev4-1}
\bibliography{zznnlo}

\begin{thebibliography}{79}
\bibitem [{mun()}]{munich}%
  \BibitemOpen
  \href@noop {} {}\bibinfo {note} {{\sc Munich} is the abbreviation of
  ``MUlti-chaNnel Integrator at Swiss (CH) precision''---an automated parton
  level NLO generator by S.~Kallweit. In preparation.}\BibitemShut {Stop}%
\end{thebibliography}
\end{document}